\documentclass[usenatbib,fleqn,letters]{mnras}
\usepackage{newtxtext,newtxmath}
\usepackage[T1]{fontenc}
\usepackage{ae,aecompl}
\usepackage[pdftex]{graphicx}
\usepackage{epstopdf}

\usepackage{amsmath}	% Advanced maths commands
\usepackage{amssymb}	% Extra maths symbols

\usepackage{ctable}
\usepackage{url}
\usepackage{xspace}
\usepackage[normalem]{ulem}

\def\msun{{\rm\,M_\odot}}

\newcommand{\kms}{\, {\rm km\, s}^{-1}}

\newcommand{\be}{\begin{equation}}
\newcommand{\ee}{\end{equation}}

\def\h2{${\rm\,H_2}$}

\title[Natal Kicks and r-process Enrichment ]{On the Impact of Neutron Star Binaries Natal-Kick Distribution on the Galactic r-process Enrichment}
\author[Safarzadeh \& C\^ot\'e]{Mohammadtaher Safarzadeh$^1$\thanks{E-mail: mts@asu.edu}, Benoit C\^ot\'e $^{2,3,4}$\\
	$^1$School of Earth and Space Exploration, Arizona State University, Tempe, AZ 85287-1404, USA;\\
	$^2$Department of Physics and Astronomy, University of Victoria, Victoria, BC, V8W 2Y2, Canada;\\
    $^3$National Superconducting Cyclotron Laboratory, Michigan State University, East Lansing, MI, 48824, USA;\\
    $^4$Joint Institute for Nuclear Astrophysics - Center for the Evolution of the Elements, USA
}

\usepackage[all]{hypcap}

\pubyear{2017}
\begin{document}

\label{firstpage}
\pagerange{\pageref{firstpage}--\pageref{lastpage}}
\maketitle

\begin{abstract} 
We study the impact of the neutron star binaries' (NSBs) natal kick distribution on the Galactic r-process enrichment. We model the growth of a Milky Way type halo based on N-body simulation results and its star formation history based on multi epoch abundance matching techniques. We consider the NSBs that merge well beyond the galaxy's effective radius ($>2\times R_\mathrm{eff}$) do not contribute to Galactic r-process enrichment.
Assuming a power-law delay-time distribution (DTD) function ($\propto t^{-1}$) with $t_\mathrm{min}=30$\,Myr for binaries' coalescence timescales, and an exponential profile for their natal kick distribution with an average value of 180\,km\,s$^{-1}$, we show that up to $\sim$\,40\% of all formed NSBs do not contribute to r-process enrichment by $z=0$, either because they merge far from the galaxy at a given redshift (up to $\sim$\,25\%) or have not yet merged by today ($\sim$\,15\%). Our result is largely insensitive to the details of the DTD function. Assuming a constant coalescence timescale of 100 Myr well approximates the adopted DTD with 30\% of the NSBs not contributing to r-process enrichment. Our results, although rather dependent on the adopted natal kick distribution, represent a first step towards estimating the impact of natal kicks and DTD functions on r-process enrichment of galaxies that would need to be incorporated in the hydrodynamical simulations. 
%\bc{157577 left unmerged out of 1084263.}
\end{abstract}


%%%%%%%%%%%%%%%%%%%%%%%%%%%%%%%%%%%%%%%%%%%%%%%%%%%%%%%%%%%%%%%%%%%%%%%%%%%%%%%%%%%%%%%%%%%%%%%

\section{Introduction}

Neutron star mergers (NSMs) and core collapse supernovae (SNcc) are the two main candidates to explain the observed Galactic r-process enrichment \citep{Cowan:1991ca,Woosley:1994ih,Rosswog:1999wz,Rosswog:2000nj,Argast:2004hg}. Though the rate of NSMs is orders of magnitude less than SNcc, they have orders of magnitude higher r-process yields. 
Although heavy r-process elements could be produced under extreme conditions of a magnetorotational core-collapse supernovae \citep{Winteler:2012fv,Nishimura:2015ki}, recent hydrodynamical simulations of core-collapse supernovae with detailed treatment of neutrino transport show that it is highly difficult to synthesize r-process elements heavier than A>110 \citep{Wanajo:2010js,MartinezPinedo:2012ev,Roberts:2012gc,Fischer:2012cl,Wanajo:2013io}. This leaves NSMs as potentially the only robust source of heavy r-process elements
\citep{Wanajo:2014ia,Goriely:2015hn}.

Hydrodynamical simulations have been successful at reproducing the observed abundance of r-process elements by assuming NSMs as the only source of r-process elements in the Galaxy \citep{vandeVoort:2015jw,Shen:2015gc}. \cite{2015ApJ...814...41H} demonstrated that, even with a delay time of 100\,Myr, NSM ejecta can still enrich low-metallicity stars when reducing the star formation efficiency of low-mass building block galaxies. This conclusion was also predicted by the simple model of \cite{2015ApJ...804L..35I} and the semi-analytic model of \cite{2016ApJ...830...76K}. We refer to \cite{2017ApJ...836..230C} for a review of recent r-process enrichment studies for the MW.     

There are two main sources of uncertainty when modeling NSMs as sources of r-process enrichment in the Galaxy. The NSM rate, which can vary by 2 or 3 orders of magnitudes according to population synthesis models \citep{2012ApJ...759...52D}, and r-process ejected mass in a NSM event with a wide range from $10^{-4} - 4\times10^{-2}$\,M$_{\odot}$ \citep{Oechslin:2002iu,Goriely:2011fa,Korobkin:2012cp,Piran:2013da}. The escape of NSBs from the host halo due to the natal kicks is the third uncertain factor when modeling r-process enrichment with NSMs. Depending on the merging time scale associated to them and the host halo mass they reside in, NSBs can merge well beyond their host galaxy \citep{1999MNRAS.305..763B,1999ApJ...526..152F,2006ApJ...648.1110B,Zemp:2009gu,Kelley:2010iq,2014ApJ...792..123B} and therefore be regarded as host-less merging events which do not contribute to r-process enrichment. 

The impact of natal kicks has been studied before in different contexts mainly related to short gamma-ray burst (sGRBs). \citet{2014ApJ...792..123B}
studied the impact of natal kicks on the fraction of host-less sGRBs \citep{Berger:2010de} from N-Body simulations and merger tree studies. 
\citet{Kelley:2010iq} studied this in the context of spatial distribution of sGRBs on the sky and the predictions for future gravitational wave experiments. \citet{Bramante:2016kp} considered the natal kick impact on r-process enrichment of ultra faint dwarf galaxies \citep[UFDs,][]{Brown:2012jo,Frebel:2012ja,Vargas:2013ei} and concluded that natal kicks most certainly remove the NS binary from the UFD progenitor at redshifts of reionization and suggested alternative pathways to explode a lone NS. Unless very short timescales for merging of a NS binary is possible due to Kozai effect \citep{Shappee:2013ea,Beniamini:2016kw} or a common envelope scenario \citep{2002ApJ...572..407B}, the NS binary will tend to travel away from the host and therefore not contribute to r-process enrichment.

Here we investigate how such considerations would affect the results of r-process enrichment codes. We follow the MW progenitor halo growth history modeled analytically based on Bolshoi N-Body simulations merger trees. Given a parametrized star formation history (SFH) for a MW type halo at $z=0$ based on multi epoch abundance matching techniques, we compute the fraction of NSBs that would merge within a multiple of their host galaxy's effective radius given a natal kick velocity probability distribution function (PDF) and a delay-time distribution assigned to the NSBs. 

In \S2 we describe our method in more detail. In \S3 we present our results. In \S4 we present a discussion of our results and in \S5 we give conclusions. 

%%%%%%%%%%%%%%%%%%%%%%%%%%%%%%%%%%%%%%%%%%%%%%%%%%%%%%%%%%%%%%%%%%%%%%%%%%%%%%%%%%%%%%%%%%%%%%%

%%%%%%%%%%%%%%%%
%%%%%%%%%%%%%%%%
\section{method}
%%%%%%%%%%%%%%%%
%%%%%%%%%%%%%%%%
In this section, we describe how we model the evolution of the dark matter halo and its central galaxy as a function of redshift. We then describe the formation and the trajectory calculation of NSBs in order to predict what fraction of them do not contribute to the galactic r-process enrichment.

%%%%%%%%%%%%%%%%%%%%%%%%%%%%%%%%%%%%%%%%%%%%%%
\subsection{Evolution of the Dark Matter Halo}
%%%%%%%%%%%%%%%%%%%%%%%%%%%%%%%%%%%%%%%%%%%%%%
\label{sect_evol_DM}
We consider a MW-like dark matter halo with a current mass of $10^{12}$\,$\msun$ \citep{2015MNRAS.453..377W}. We use the relations derived by \cite{2013ApJ...770...57B} from the \textit{Bolshoi} simulation \citep{2011ApJ...740..102K} to calculate the evolution of the dark matter mass $M_\mathrm{DM}$ as a function of redshift ($z$).
We express the halo radius ($R_{200}$) defined as enclosing overdensity of 200 times the critical density of the universe at a given redshift, $\rho_{\mathrm{cr}}(z)$,
so $M_{200}= (4\pi/3) 200 \rho_{\mathrm{cr}}(z)R_{200}^3$. With this setup, R$_{200}=209$\,kpc at $z=0$. Values for the cosmological parameters are taken from \cite{2016A&A...594A..13P}: $H_0=100 h$~km\,s$^{-1}$\,Mpc$^{-1}$, $h=0.678$, $\Omega_0=0.308$, $\Omega_{\Lambda,0}=0.692$.

%%%%%%%%%%%%%%%%%%%%%%%%%%%%%%%%%%%%%%%%%%%%%%%%
\subsection{Dark Matter Gravitational Potential}
%%%%%%%%%%%%%%%%%%%%%%%%%%%%%%%%%%%%%%%%%%%%%%%%
At each redshift, the radial gravitational potential profile of the NFW dark matter halo is 

\begin{equation}
\Phi(r)=-\frac{GM_\mathrm{DM}\mathrm{ln}(1+c_{200}r/R_{200})}{\left[\mathrm{ln}(1+c_{200})-c_{200}/(1+c_{200})\right]r},
\end{equation}

\noindent where $G$ is the Newton gravitational constant and $c_{200}$ is the concentration parameter of the dark matter halo we assign following \cite{2014MNRAS.441.3359D}. The acceleration $g$ acting on a NS binary as a function of radius is given by the derivative of the potential profile,

\begin{equation}
\label{eq_g}
g(r) = \frac{d\Phi}{dr}=A_\Phi \left(\frac{\beta}{r(1+\beta r)} - \frac{\mathrm{ln}(1+\beta r)}{r^2}\right),
\end{equation}

\begin{equation}
\beta\equiv c_{200}/R_{200},
\end{equation}

\begin{equation}
A_\Phi=-\frac{GM_\mathrm{DM}}{\mathrm{ln}(1+c_{200})-c_{200}/(1+c_{200})}.
\end{equation}
In these last equations, $M_\mathrm{DM}$, $R_{200}$, and $c_{200}$ are constant with radius, but are still dependent on redshift.

%%%%%%%%%%%%%%%%%%%%%%%%%%%%%%%%%%%%%%%%%%%%
\subsection{Evolution of the Central Galaxy}
%%%%%%%%%%%%%%%%%%%%%%%%%%%%%%%%%%%%%%%%%%%%
\label{sect_evol_gal}
We describe the central galaxy by its effective radius and its stellar mass content. For star formation history of a MW type halo, we adopt the parametrization presented in \cite{2013MNRAS.428.3121M} which is based on a multi epoch abundance matching technique. From this relation, we create several redshift bins with constant redshift intervals, convert redshifts into times, and integrate the star formation history to recover the total stellar mass of the galaxy for each redshift bin. In our model, at $z=0$, the star formation rate is 2.63\,M$_\odot$\,yr$^{-1}$ and the total integrated stellar mass is $5.43\times10^{10}$\,M$_\odot$.

We set the effective radius $R_\mathrm{eff}$ of the galaxy to a fraction of the virial radius of its host dark matter halo \citep{2013ApJ...764L..31K},

\begin{equation}
\label{eq_r_eff}
R_\mathrm{eff}(z)\equiv0.032R_{200}(z).
\end{equation}

\noindent The proportionality constant has been tuned to reproduce the observed relation derived by \cite{2014ApJ...788...28V} between $R_\mathrm{eff}$, redshift, and the stellar mass of late-type blue galaxies up to $z=3$. This value is about twice the value derived by \cite{2013ApJ...764L..31K} but is still within the observational scatter (see their Figure 1). At $z=0$, Equation~\ref{eq_r_eff} yields $R_\mathrm{eff}\approx 6.7$\,kpc. 

%%%%%%%%%%%%%%%%%%%%%%%%%%%%%%%%%%%
\subsection{Neutron Star Binaries}
%%%%%%%%%%%%%%%%%%%%%%%%%%%%%%%%%%%
\label{sect_evol_NS}
In this work, we assume that $2\times10^{-5}$ NSB system will lead to a NSM per units of stellar mass formed, a number adopted in the chemical evolution study of \cite{2017ApJ...836..230C}.  The input number of NSMs is typically calibrated to reproduce the current Galactic NSM rate estimated by binary pulsars (\citealt{2004ApJ...614L.137K,2010CQGra..27q3001A}), as in the chemical evolution studies of \cite{2014MNRAS.438.2177M}, \cite{2015A&A...577A.139C}, \cite{2015ApJ...814...41H}, \cite{vandeVoort:2015jw}, and \cite{2015MNRAS.452.1970W}. This number can also be extracted from population synthesis predictions (\citealt{2012ApJ...759...52D}), as in \cite{2015ApJ...804L..35I}, or derived from NSM yields and observed chemical abundances (\citealt{Shen:2015gc}). In our study, however, the total number of NSMs is only relevant for the statistics of our Monte Carlo calculations and does not have an impact on our conclusions (see Section~\ref{sect_dis}).

Each NS is expected to receive a natal kick that can reach several hundreds of km\,s$^{-1}$. Different probability distributions as been proposed to explain the observed NS velocities such as a single Maxwellian distribution (e.g., \citealt{1997MNRAS.291..569H,2005MNRAS.360..974H}) and a bimodal distribution (e.g., \citealt{1998ApJ...496..333F,2002ApJ...568..289A}). However, the final velocity kick imparted to a NSB system is more complicated, as the system experiences two different kicks and the orbital properties of the binary must be taken into account \citep{1998ApJ...496..333F}. \cite{2002ApJ...572..407B} showed that the velocity of a NSB significantly changes after the second kick generated by the second supernova explosion. In addition, a too large velocity kick imparted to a NS can unbound or significantly increase the coalescence timescale of NSBs. (\citealt{1999A&A...346...91B,2000ApJ...530..890K}).

We assign a unique initial natal kick velocity $v_\mathrm{kick}$ to each NSB by randomly sampling an exponential probability distribution function (PDF) defined as \citep{2014ApJ...792..123B}
\begin{equation}
\mathrm{PDF}(v_\mathrm{kick})=\mathrm{exp}\left(-\frac{v_\mathrm{kick}}{<v>}\right),
\end{equation}
where $<v>$ represents the average natal kick velocity, which is set to 180\,km\,s$^{-1}$. This PDF has been derived to approximate the NS binary natal kick distribution predicted by the population synthesis model of \cite{1998ApJ...496..333F}, which uses a bimodal kick distribution for each individual NS.
%In a forthcoming paper, we plan to repeat our methodology by directly sampling the NS binary kick PDF predicted by different population synthesis models \citep{1998ApJ...496..333F,2002ApJ...572..407B}.}

Once a kick is sampled, a unique NSM coalescence time $t_\mathrm{coal}$ is assigned to each NSB by randomly sampling a power-law delay-time distribution (DTD) function defined as (e.g., \citealt{2012ApJ...759...52D})
\begin{equation}
\mathrm{PDF}(t_\mathrm{coal})=t^{-1}_\mathrm{coal}.
\end{equation}
For our fiducial case, we assume that the minimum and maximum coalescence times are 30\,Myr and 10\,Gyr, respectively. The minimum time is motivated by the standard population synthesis models of  \cite{2016ApJ...819..108B} (see Figure~8 in \citealt{2017ApJ...836..230C}). We note that some NSBs can have a coalescence time larger than the Hubble time (\citealt{2008LRR....11....8L}). However, because of the functional form of the adopted DTD, extending the maximum coalescence time beyond 10\,Gyr will have negligible impact on our results.

To perform our Monte Carlo simulation in a consistent manner, we restructure the redshift bins to ensure to form the same number (here 1000) of NSM candidate binaries in each bin.  Otherwise, the scatter in the results would vary as a function of redshift, which would be a numerical artifact. In Section~\ref{sect_results}, we compare our results with constant coalescence timescales of 30 and 100\,Myr to connect with some galactic r-process enrichment studies (e.g., \citealt{2014MNRAS.438.2177M, 2015A&A...577A.139C, 2015MNRAS.452.1970W, 2017ApJ...836..230C, 2017MNRAS.466.2474H}).

%\begin{figure}
%\includegraphics[width=3.25in]{cumul_f_NS_2_r_eff.pdf}
%\caption{Cumulated fraction of NSBs, formed at %redshift $z$, that do not contribute to the chemical %evolution by $z=0 $, for 100 simulations. Lines are the %same as in Figure~\ref{fig_f_NS_5_r_eff}.}
%\label{cumul_fig_f_NS_5_r_eff}
%\end{figure}

All PDFs are assumed to be similar at all redshifts. This is a first order approximation, since according to population synthesis models, the number and DTD function of NSMs should vary with metallicity and therefore with redshift (e.g., \citealt{2012ApJ...759...52D,2016ApJ...819..108B}).

%%%%%%%%%%%%%%%%%%%%%%%%%%%%%%%%%%%%%%%%%%%%%%%%%%%
\subsection{Non-Contributing Neutron Star Binaries}
%%%%%%%%%%%%%%%%%%%%%%%%%%%%%%%%%%%%%%%%%%%%%%%%%%%
\label{sect_non_contr}
Here we describe our procedure to calculate the fraction of NSBs that do not contribute to r-process enrichment, either because they merge beyond the galaxy or because they do not have time to merge by $z=0$. Throughout this paper, all of our NSBs have a coalescence time shorter than 10\,Gyr (see Section~\ref{sect_evol_NS}).

For each redshift bin, starting at the highest redshift, we define an enclosing radius $R_\mathrm{enc}$ based on the effective radius of the galaxy,
\begin{equation}
\label{eq_R_esc}
R_\mathrm{enc}\equiv f_\mathrm{enc}R_\mathrm{eff}.
\end{equation}
All NSMs that occur above this threshold radius are assumed to not contribute to r-process enrichment of the galaxy. 
NSM ejecta have very large speeds around $v_{\rm ejc}\sim 0.2 c$ with a stopping length of $l_s \sim (2.6/n_{\rm H})$\,kpc where $n_{\rm H}$ is the number density of the Hydrogen atom in cm$^{-3}$ in which the NSM goes off \citep{2016ApJ...830...76K}. If the NSM event occurs outside the enclosing radius that we consider in this paper, the medium has densities of $n_{\rm H}\ll0.1$~cm$^{-3}$, which means $l_s \gg 30$\,kpc. Such large stopping length translates into a very diluted gas in r-process elements. Stars forming out of such diluted gas will show no detectable r-process enhancement, and even if this gas
falls back into the ISM of the host galaxy, its contribution to r-process enrichment compared to the gas that is enriched with a NSM going off inside the enclosing radius is deemed negligible.

In Section~\ref{sect_results}, we test $f_\mathrm{enc}=2$ and 4 with a fiducial value of 2. Although we assume that all NSBs originate from the center of the host dark matter halo, we do not explore values lower than $f_\mathrm{enc}=2$ because it represents the minimum length to cross the galaxy's effective volume from one side to the other. In our calculation, we do not account for a stellar disc geometry and potential. 
%and for the possible re-accretion of escaped r-process ejecta onto the galaxy by fallback (see Section~\ref{sect_dis} for a discussion).

For each NSB, we randomly sample $v_\mathrm{kick}$ and $t_\mathrm{coal}$ and calculate the trajectory of the binary system by integrating the classical equations of motion. With an initial velocity of $v_\mathrm{kick}$, we follow the trajectories until $t_\mathrm{coal}$ using the radius-dependent gravitational acceleration defined in Equation~\ref{eq_g}, which also depends on the characteristics of the dark matter halo (see Section~\ref{sect_evol_DM}). We assume that the radial gravitational potential profile does not change through a trajectory calculation.

Depending on $v_\mathrm{kick}$, $t_\mathrm{coal}$, and the gravitational potential, the trajectory can be oscillating around the galaxy center. After a time $t_\mathrm{coal}$, we compare the final radius with $R_\mathrm{enc}$ (Equation~\ref{eq_R_esc}) and define whether or not the NSB will contribute to the galactic enrichment of r-process elements. If the radius becomes larger than $R_\mathrm{200}$ during the calculation, we stop to follow the trajectory and assume that the NSB do not contribute to  r-process enrichment.

%%%%%%%%%%%%%%%%%
%%%%%%%%%%%%%%%%%
\section{results}
\begin{figure}
\includegraphics[width=3.25in]{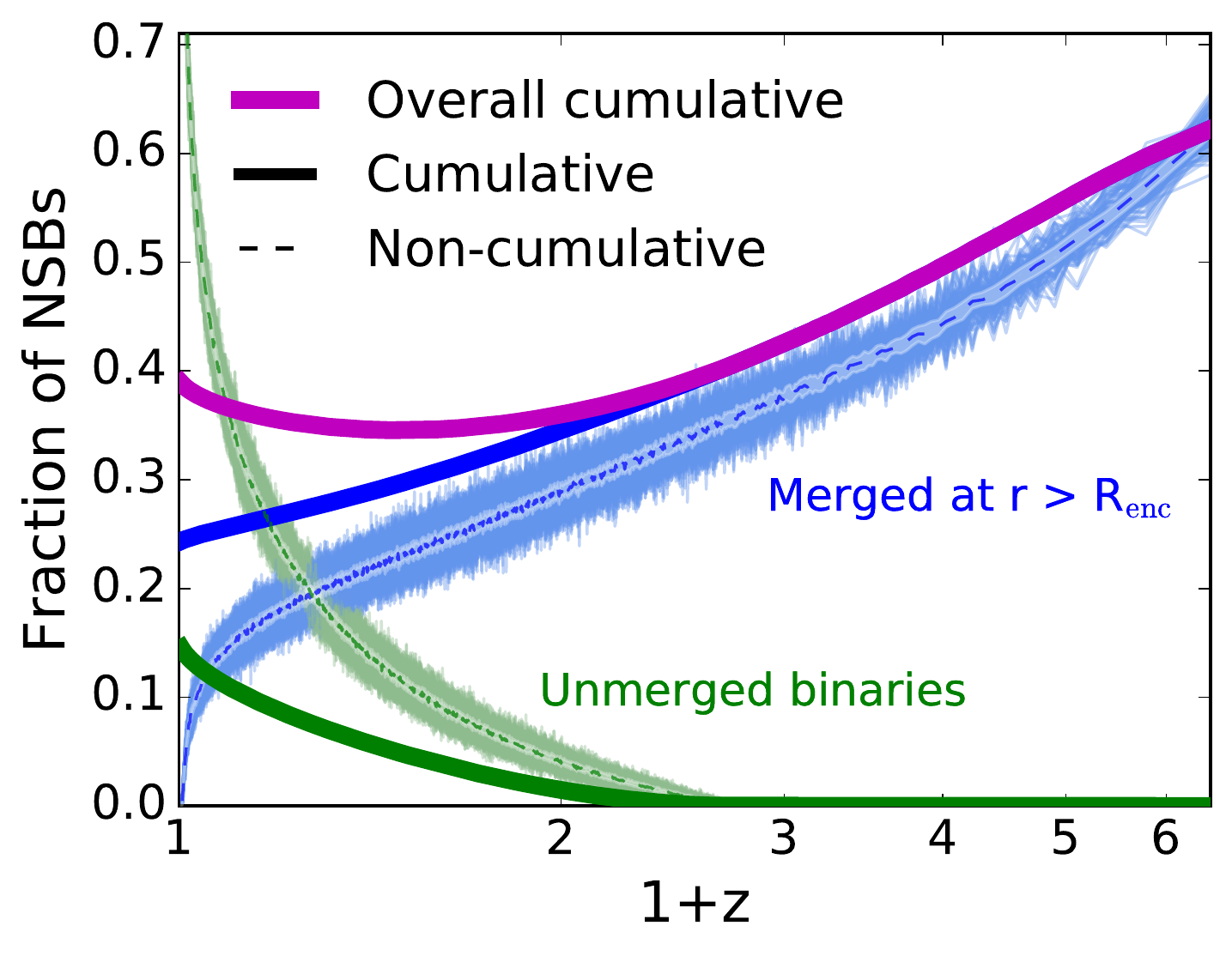}
\caption{Fraction of NSBs that do $\emph{not}$ contribute to r-process enrichment. 
The blue dashed line represents the  fraction of NSBs formed at redshift $z$ that do not contribute to r-process enrichment at the time of their merging because this occurs far from the galaxy ($r>2R_\mathrm{eff}$). The green dashed line shows the fraction of NSBs formed at redshift $z$ that do not merge by $z=0$ because of the delay in their coalescence. Shaded light blue and light green regions show the results of 100 simulations. The smaller white shaded areas wrapping the dashed lines represent the 68\,\% confidence interval. The solid lines show the cumulative results, meaning the fraction of all the NSBs formed by redshift $z$ have not contributed to r-process enrichment either because they merge outside a multitude of the galaxy's effective radius (solid blue) or have not merged yet by redshift $z$ (solid green). The purple solid line shows the overall cumulative non-contributing fraction of NSBs by redshift $z$ which is the sum of the green and blue solid lines.}
\label{fig_f_NS_5_r_eff}
\end{figure}

%%%%%%%%%%%%%%%%%
%%%%%%%%%%%%%%%%%
\label{sect_results}
Figure~\ref{fig_f_NS_5_r_eff} shows our main results for a MW type halo where we have tracked the NSBs trajectory inside their host virialized halo. The blue dashed line shows the median of 100 realizations of our simulations for
the fraction of NSBs formed at a given redshift $z$ that merge beyond the Galaxy's effective radius and therefore considered as not contributing to r-process enrichment of the halo at $z=0$. The green dashed line shows NSBs formed at $z=z$ and have not yet merged by $z=0$ because of their delayed coalescence timescale.
As the potential well of the host halo becomes deeper with time, the fraction of NSBs that merge outside the Galaxy and do not contribute to r-process enrichment drops with time from $\sim$60\,\% at $z=5$ to almost zero percent at $z=0$. We have assumed a delay-time distribution in the form of a power law from $t_\mathrm{min}=30$\,Myr to $t_\mathrm{max}=10$\,Gyr for our fiducial analysis. At redshifts lower than $z\sim2$ the NSBs start to not contribute to r-process enrichment because of the delay they experience before merging (green dashed line). This fact will increase the total fraction of the NSBs that eventually not contribute to r-process enrichment of the Galaxy at $z=0$. 

The cumulative picture of the results described above is shown with solid lines in Figure~\ref{fig_f_NS_5_r_eff}. The solid blue line shows the cumulative fraction of the NSBs that never contribute to r-process enrichment by $z=z$ due to merging beyond $R_\mathrm{enc}$ while the solid green line shows the cumulative fraction of NSBs, formed by $z=z$, that do not merge by $z=0$. In our fiducial case, about 40\% of the NSBs that do not contribute to r-process enrichment by $z=0$ is because of the delay time they experience before merging. The solid purple line shows the sum of the two cumulative solid blue and solid green lines and therefore represents the overall cumulative fraction of all NSBs that do not contribute to r-process enrichment as a function of redshift.

\begin{figure}
\includegraphics[width=3.25in]{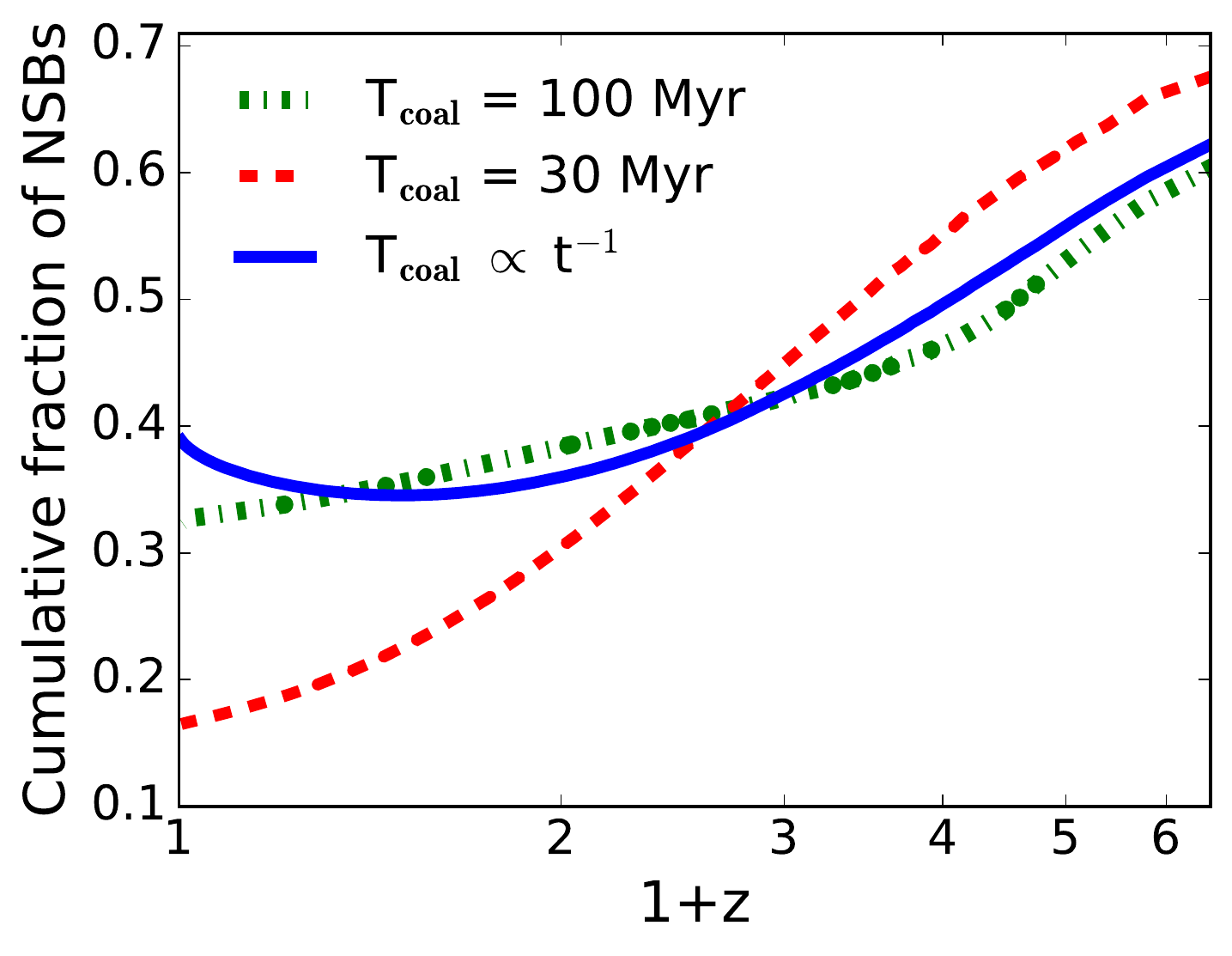}
\caption{Impact of the delay-time distribution assumption of NSMs on the cumulated fraction of NSBs formed by redshift $z$, that do not contribute to r-process evolution as a function of redshift. Each line represents the median value of 100 simulations. The blue solid line shows the result of using a power-law delay-time distribution function with a minimum coalescence time of 30\,Myr. The red dashed and green dot-dashed} lines are the result of using a constant coalescence timescale of 30 and 100\,Myr, respectively, after which all NSMs occur in a simple stellar population. In these two last cases, no NSM is left unmerged by $z=0$ because there is no NSM with long delay times.
\label{fig_impact_DTD_cte}
\end{figure}

In Figure~\ref{fig_impact_DTD_cte}, we compare the overall cumulative fraction of not-contributing NSBs with different assumptions regarding the coalescence timescale of the binaries. The red dashed and green dot-dashed lines show the cumulative fraction of NSBs that do not contribute to r-process enrichment by $z=0$ assuming a constant $t_\mathrm{coal}$ of 30 and 100\,Myr, respectively. The short coalescence timescales in the case of constant $t_\mathrm{coal}$ leave no room for the NSBs to not merge by $z=0$. Using a constant timescale instead of a DTD for the coalescence times of NSMs decreases the total fraction of non-contributing NSBs from $\sim40\%$ to $\sim30\%$ for $t_\mathrm{coal}=100$\,Myr, and to $\sim15\%$ in the case of $t_\mathrm{coal}=30$\,Myr.

We also changed the minimum timescale for merging in DTD assumption from 10\,Myr to 100\,Myr and just observed a difference of 10\% in the total fraction of the NSBs compared to the fiducial case. This is because longer timescale for merging results in both an increase in the fraction of NSBs that do not have time to merge and also let the NSBs to travel further out inside the virialized halo, an effect that is in place since $z\sim 1.5$. 

\begin{figure}
\includegraphics[width=3.25in]{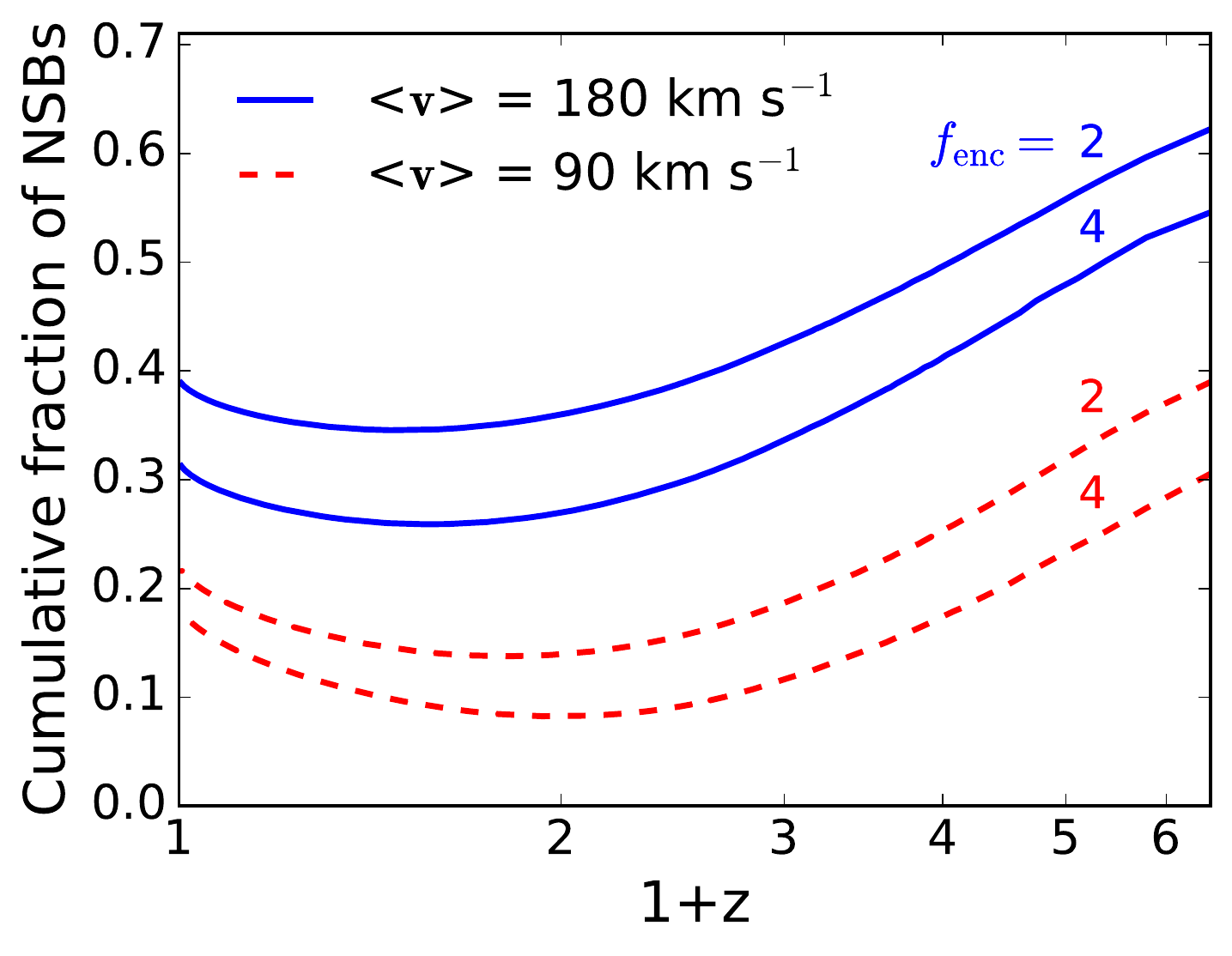}
\caption{Impact of the enclosing radius on the cumulative fraction of NSBs that do not contribute to r-process enrichment by redshift $z$. Each line represents the median value of 100 simulations. Different lines represent different enclosing radius $R_\mathrm{enc}=f_\mathrm{enc}R_\mathrm{eff}$, as indicated next to each line in the panel. Solid blue (dashed red) lines present the results when adopting a natal kick distribution with $<v>=180~(90)\kms$.}
\label{fig_impact_f_esc}
\end{figure}

In Figure~\ref{fig_impact_f_esc} we compare the assumption we put for the radius beyond which a NSM do not contribute to r-process enrichment. We compare two cases where $R_\mathrm{enc}=2$ and 4~$\times\,R_\mathrm{eff}$ and 
see that considering $f_{\rm enc}=4$ leads to 25\% more NSBs contributing to r-process enrichment (blue solid lines). The effect is more pronounced when using higher median natal kick velocities. With a low median velocity (red dashed lines), NSBs at low redshift eventually become unable to travel beyond the defined enclosed radius, regardless of their coalescence times. In that case, by increasing $f_{\rm enc}$, the cumulative fraction of non-contributing NSBs approaches 15\,\%, the fraction of NBSs that do not have time to merge by $z=0$.

%\begin{figure}
%\includegraphics[width=3.25in]{impact_t_min.pdf}
%\caption{Impact of the coalescence timescale on the cumulated fraction of NSBs, formed at redshift $z$, that does not contribute to the chemical evolution as a function of redshift. Each line represents the median value of 100 simulations assuming $R_\mathrm{esc}=5R_\mathrm{eff}$. Upper panel: Impact of using different minimum coalescence times (different colors) assuming a power-law DTD function. Lower panel: Impact of using different constant coalescence times (different colors) without using any DTD function.}
%\label{fig_impact_t_min}
%\end{figure}

%\begin{figure}
%\includegraphics[width=3.25in]{unmerged.pdf}
%\caption{Impact of the minimum coalescence timescale (different colors) on the cumulated fraction of NSBs that do not have time to merge before redshift $z$. Each line represents the median value of 100 simulations assuming $R_\mathrm{esc}=5R_\mathrm{eff}$ and a power-law DTD function.}
%\label{fig_impact_t_min}
%\end{figure}

%\begin{figure}
%\begin{center}
%\includegraphics[width=3.6in]{impact_v_mean.png}
%\caption{Median cumulated fraction of NS-NSBs that does not contribute to the chemical evolution as a function of redshift, for 50 simulations. Different colors represent different mean natal kick velocities (see Section~\ref{sect_evol_NS}).}
%\label{fig_impact_v_mean}
%\end{center}
%\end{figure}

\section{discussion}
\label{sect_dis}
This work consists of a first step towards a more complex and realistic approach we will present in an upcoming work. We did not consider a disc shape for the galaxy at low redshifts when tracing the trajectory of the binary. Depending on the natal kick and coalescence timescale of a NSB, the NSM will have more chance to occur beyond the star-forming region if the binary system is kicked perpendicularly to the galactic plane. For the largest enclosing radius considered in this work ($4~R_\mathrm{eff}$), the orientation of the kick relative to the galactic plane becomes less problematic. In addition, we assumed that all the trajectories originate from the center of the galaxy. A more realistic approach would be to spatially distribute the initial position of the binaries according to the stellar density profile, which could be done when accounting for the shape of the central galaxy.

Introducing a disc geometry in our work would also allow to account for the impact of binaries angular momentum on their trajectory. Depending on the angle between the natal kick velocity vector and that of its angular momentum, the effect of the natal kick will be different, which could modify the predicted fraction of non-contributing NSMs on r-process enrichment. One of our next improvements will be to follow the trajectory of NSBs in 3D instead of 1D and to account for their location within the disc along with the orientation of their kick with respect to the galactic plane when the disk potential is also taken into account. 
%In addition, we did not account for the fact that some of the mass ejected by NSM outside the enclosing radius could be reintroduced in the galaxy by fallback. This effect should be more important when NSMs occur relatively near the galactic plane. Such recycling is difficult to predict with our simple approach but can be done with hydrodynamic simulations.

%\begin{figure}
%\includegraphics[width=3.25in]{impact_f_esc_90kms.pdf}
%\caption{Same as in Figure~\ref{fig_impact_f_esc}, but using an average natal kick velocity of $<v>=90$\,km\,s$^{-1}$ instead of 180\,km\,s$^{-1}$.}
%\label{fig_impact_f_esc_90kms}
%\end{figure}

Our fiducial model is based on the assumption that the natal kick PDF of NSBs follows an exponential distribution with an average velocity of $<v>=180$\,km\,s$^{-1}$ (see Section~\ref{sect_evol_NS}). \cite{2013ApJ...776...18F}, however, derived lower velocities in the range of 20-140\,km\,s$^{-1}$ with a median value around 60\,km\,s$^{-1}$. In addition, \cite{2016MNRAS.456.4089B} showed that the natal kick distribution can be bimodal with a low-velocity component below $\sim30$\,km\,s$^{-1}$. Following \cite{2014ApJ...792..123B}, we tested an exponential velocity PDF with $<v>=90$\,km\,s$^{-1}$ to represent \cite{2013ApJ...776...18F} findings. As shown in Figure~\ref{fig_impact_f_esc}, reducing $<v>$ by a factor of 2 reduces the fraction of non-contributing NSBs by a factor between 1.5 and 2, which demonstrates that the choice of the natal kick PDF has a non-negligble impact on our results.

The rates of NSMs derived from pulsar luminosity distribution (\citealt{2004ApJ...614L.137K,2010CQGra..27q3001A}) and from gravitational wave measurements (Advanced LIGO, \citealt{2016ApJ...832L..21A}) are uncertain by about 3 orders of magnitude. A similar range is predicted by population synthesis models (\citealt{2012ApJ...759...52D}). However, as we do not vary the number of NSMs (per units of stellar mass formed) throughout our calculations and only calculate the $\emph{fraction}$ of non-contributing NSBs, the total number of NSMs formed in our simulation does not impact our results and therefore these large uncertainties do not affect the quantities derived in this paper. That said, those uncertainties will affect the contribution of NSMs on the evolution of r-process in the MW, as the level of enrichment is directly proportional to the number of enriching sources.

\section{conclusions}
\label{sect_con}
We presented a parametrized approach to quantify the impact of NSB natal kicks on r-process enrichment of a MW-like galaxy by NSMs. The progenitor halo mass is parametrized to be consistent with the results of N-body simulations. The SFH is adopted from an abundance matching technique and the size evolution of the galaxy is taken to be in agreement with observations. NSMs are born with a natal kick and delay time for merging and we follow their trajectory in radial direction inside their host halo potential. If a NSB merges beyond twice the galaxy's effective radius, we consider that binary to not contribute to r-process enrichment.

Given the caveats stated in Section~\ref{sect_dis} and the absence of gas recycling in our model, we predict that up to 40\% of all the NSBs formed in the entire star formation history of a MW-like galaxy do not contribute to r-process enrichment either because they merge well beyond the galaxy's effective radius at a given redshift or because of the delay they experience in merging. About 15\% of all NSBs are predicted to be free floating in the MW halo. This result is based on a power law DTD motivated by \cite{2012ApJ...759...52D} and an exponential natal kick velocity suggested by \cite{2014ApJ...792..123B} with an average velocity of 180 $\kms$. 

Implementing the natal kicks together with delay-time distributions in hydrodynamical simulations of r-process enrichment such as those carried out by \citet{vandeVoort:2015jw}, \cite{2015ApJ...814...41H}, and Safarzadeh \& Scannapieco (in prep) would be the closest simulation to reality, but it is not clear
how futuristic this approach would be in order to arrive at a significant statistics. Our results in this paper is a first order estimate on the impact of the natal kicks and provides correction factors for r-process enrichment as a function of redshift for a MW type halo, though the corrections would be more in agreement with hydro simulations at higher ($z$>1) redshifts as the geometry of a disc is not taken into account in this work.

In a forthcoming paper, we plan to use the NSM delay-time and natal kick PDFs predicted by population synthesis models (e.g., \citealt{1998ApJ...496..333F,2002ApJ...572..407B,2012ApJ...759...52D,2016ApJ...819..108B}). This will enable to account for the variation with metallicity (redshift) of the total number of NSMs per unit of stellar mass formed as well as the variation of the shape of the PDFs. Such variations are rarely included in galactic chemical evolution simulations (but see \citealt{2016A&A...589A..64M}), although they can significantly modify the chemical evolution trends \citep{2017ApJ...836..230C}. This complementary study will provide insights into the impact of using different modeling assumptions for NSMs.

\section{Acknowledgements}

We are thankful to the anonymous referee for their constructive comments. We are also thankful to Chris Fryer, Chris Belczynski, and Evan Scannapieco for useful discussions. 
MS and BC are supported by the National Science Foundation (USA) under Grant No. PHY-1430152 (JINA Center for the Evolution of the Elements).

%\ms{comments from Smadar:
%\\
%The natal kick values are low compared with what they use in their research. 
%read Kalogera et al 1998,Kalogera 2000, Hurley et al 2002, Belczynski et al 2006
%cite Lu \& Naoz in prep.
%Hansen \& Phinney 1997
%Hobbs et al 2004,2005
%Smadar et al consider a pdf for natal kick that is a normal distribution with mean value of 400 km/sec and std = 265 km/s. Consistent with Hobbs et al. Arzoumanian et al 2002
%\\
%\\
%Comments from Evan:
%\\
%The R200 formula should note have the Omega dependence, or we should make sure it is correct. Also we have not justified well enough why we chose an exponential pdf.
%}
\bibliographystyle{mnras}
\bibliography{references,the_entire_lib}

\begin{thebibliography}{}
\makeatletter
\relax
\def\mn@urlcharsother{\let\do\@makeother \do\$\do\&\do\#\do\^\do\_\do\%\do\~}
\def\mn@doi{\begingroup\mn@urlcharsother \@ifnextchar [ {\mn@doi@}
  {\mn@doi@[]}}
\def\mn@doi@[#1]#2{\def\@tempa{#1}\ifx\@tempa\@empty \href
  {http://dx.doi.org/#2} {doi:#2}\else \href {http://dx.doi.org/#2} {#1}\fi
  \endgroup}
\def\mn@eprint#1#2{\mn@eprint@#1:#2::\@nil}
\def\mn@eprint@arXiv#1{\href {http://arxiv.org/abs/#1} {{\tt arXiv:#1}}}
\def\mn@eprint@dblp#1{\href {http://dblp.uni-trier.de/rec/bibtex/#1.xml}
  {dblp:#1}}
\def\mn@eprint@#1:#2:#3:#4\@nil{\def\@tempa {#1}\def\@tempb {#2}\def\@tempc
  {#3}\ifx \@tempc \@empty \let \@tempc \@tempb \let \@tempb \@tempa \fi \ifx
  \@tempb \@empty \def\@tempb {arXiv}\fi \@ifundefined
  {mn@eprint@\@tempb}{\@tempb:\@tempc}{\expandafter \expandafter \csname
  mn@eprint@\@tempb\endcsname \expandafter{\@tempc}}}

\bibitem[\protect\citeauthoryear{{Abadie} et~al.,}{{Abadie}
  et~al.}{2010}]{2010CQGra..27q3001A}
{Abadie} J.,  et~al., 2010, \mn@doi [Classical and Quantum Gravity]
  {10.1088/0264-9381/27/17/173001}, \href
  {http://adsabs.harvard.edu/abs/2010CQGra..27q3001A} {27, 173001}

\bibitem[\protect\citeauthoryear{{Abbott} et~al.,}{{Abbott}
  et~al.}{2016}]{2016ApJ...832L..21A}
{Abbott} B.~P.,  et~al., 2016, \mn@doi [\apjl] {10.3847/2041-8205/832/2/L21},
  \href {http://adsabs.harvard.edu/abs/2016ApJ...832L..21A} {832, L21}

\bibitem[\protect\citeauthoryear{Argast, Samland, Thielemann  \& Qian}{Argast
  et~al.}{2004}]{Argast:2004hg}
Argast D.,  Samland M.,  Thielemann F.~K.,   Qian Y.~Z.,  2004, Astronomy {\&}
  Astrophysics, 416, 997

\bibitem[\protect\citeauthoryear{{Arzoumanian}, {Chernoff}  \&
  {Cordes}}{{Arzoumanian} et~al.}{2002}]{2002ApJ...568..289A}
{Arzoumanian} Z.,  {Chernoff} D.~F.,   {Cordes} J.~M.,  2002, \mn@doi [\apj]
  {10.1086/338805}, \href {http://adsabs.harvard.edu/abs/2002ApJ...568..289A}
  {568, 289}

\bibitem[\protect\citeauthoryear{{Behroozi}, {Wechsler}  \&
  {Conroy}}{{Behroozi} et~al.}{2013}]{2013ApJ...770...57B}
{Behroozi} P.~S.,  {Wechsler} R.~H.,   {Conroy} C.,  2013, \mn@doi [\apj]
  {10.1088/0004-637X/770/1/57}, \href
  {http://adsabs.harvard.edu/abs/2013ApJ...770...57B} {770, 57}

\bibitem[\protect\citeauthoryear{{Behroozi}, {Ramirez-Ruiz}  \&
  {Fryer}}{{Behroozi} et~al.}{2014}]{2014ApJ...792..123B}
{Behroozi} P.~S.,  {Ramirez-Ruiz} E.,   {Fryer} C.~L.,  2014, \mn@doi [\apj]
  {10.1088/0004-637X/792/2/123}, \href
  {http://adsabs.harvard.edu/abs/2014ApJ...792..123B} {792, 123}

\bibitem[\protect\citeauthoryear{{Belczynski} \& {Bulik}}{{Belczynski} \&
  {Bulik}}{1999}]{1999A&A...346...91B}
{Belczynski} K.,  {Bulik} T.,  1999, \aap, \href
  {http://adsabs.harvard.edu/abs/1999A%26A...346...91B} {346, 91}

\bibitem[\protect\citeauthoryear{{Belczynski}, {Kalogera}  \&
  {Bulik}}{{Belczynski} et~al.}{2002}]{2002ApJ...572..407B}
{Belczynski} K.,  {Kalogera} V.,   {Bulik} T.,  2002, \mn@doi [\apj]
  {10.1086/340304}, \href {http://adsabs.harvard.edu/abs/2002ApJ...572..407B}
  {572, 407}

\bibitem[\protect\citeauthoryear{{Belczynski}, {Perna}, {Bulik}, {Kalogera},
  {Ivanova}  \& {Lamb}}{{Belczynski} et~al.}{2006}]{2006ApJ...648.1110B}
{Belczynski} K.,  {Perna} R.,  {Bulik} T.,  {Kalogera} V.,  {Ivanova} N.,
  {Lamb} D.~Q.,  2006, \mn@doi [\apj] {10.1086/505169}, \href
  {http://adsabs.harvard.edu/abs/2006ApJ...648.1110B} {648, 1110}

\bibitem[\protect\citeauthoryear{{Belczynski}, {Repetto}, {Holz},
  {O'Shaughnessy}, {Bulik}, {Berti}, {Fryer}  \& {Dominik}}{{Belczynski}
  et~al.}{2016}]{2016ApJ...819..108B}
{Belczynski} K.,  {Repetto} S.,  {Holz} D.~E.,  {O'Shaughnessy} R.,  {Bulik}
  T.,  {Berti} E.,  {Fryer} C.,   {Dominik} M.,  2016, \mn@doi [\apj]
  {10.3847/0004-637X/819/2/108}, \href
  {http://adsabs.harvard.edu/abs/2016ApJ...819..108B} {819, 108}

\bibitem[\protect\citeauthoryear{{Beniamini} \& {Piran}}{{Beniamini} \&
  {Piran}}{2016}]{2016MNRAS.456.4089B}
{Beniamini} P.,  {Piran} T.,  2016, \mn@doi [\mnras] {10.1093/mnras/stv2903},
  \href {http://adsabs.harvard.edu/abs/2016MNRAS.456.4089B} {456, 4089}

\bibitem[\protect\citeauthoryear{Beniamini, Hotokezaka  \& Piran}{Beniamini
  et~al.}{2016}]{Beniamini:2016kw}
Beniamini P.,  Hotokezaka K.,   Piran T.,  2016, The Astrophysical Journal
  Letters, 829, L13

\bibitem[\protect\citeauthoryear{Berger}{Berger}{2010}]{Berger:2010de}
Berger E.,  2010, The Astrophysical Journal, 722, 1946

\bibitem[\protect\citeauthoryear{{Bloom}, {Sigurdsson}  \& {Pols}}{{Bloom}
  et~al.}{1999}]{1999MNRAS.305..763B}
{Bloom} J.~S.,  {Sigurdsson} S.,   {Pols} O.~R.,  1999, \mn@doi [\mnras]
  {10.1046/j.1365-8711.1999.02437.x}, \href
  {http://adsabs.harvard.edu/abs/1999MNRAS.305..763B} {305, 763}

\bibitem[\protect\citeauthoryear{Bramante \& Linden}{Bramante \&
  Linden}{2016}]{Bramante:2016kp}
Bramante J.,  Linden T.,  2016, The Astrophysical Journal, 826, 57

\bibitem[\protect\citeauthoryear{Brown et~al.,}{Brown
  et~al.}{2012}]{Brown:2012jo}
Brown T.~M.,  et~al., 2012, The Astrophysical Journal Letters, 753, L21

\bibitem[\protect\citeauthoryear{{Cescutti}, {Romano}, {Matteucci}, {Chiappini}
   \& {Hirschi}}{{Cescutti} et~al.}{2015}]{2015A&A...577A.139C}
{Cescutti} G.,  {Romano} D.,  {Matteucci} F.,  {Chiappini} C.,   {Hirschi} R.,
  2015, \mn@doi [\aap] {10.1051/0004-6361/201525698}, \href
  {http://adsabs.harvard.edu/abs/2015A%26A...577A.139C} {577, A139}

\bibitem[\protect\citeauthoryear{{C{\^o}t{\'e}}, {Belczynski}, {Fryer},
  {Ritter}, {Paul}, {Wehmeyer}  \& {O'Shea}}{{C{\^o}t{\'e}}
  et~al.}{2017}]{2017ApJ...836..230C}
{C{\^o}t{\'e}} B.,  {Belczynski} K.,  {Fryer} C.~L.,  {Ritter} C.,  {Paul} A.,
  {Wehmeyer} B.,   {O'Shea} B.~W.,  2017, \mn@doi [\apj]
  {10.3847/1538-4357/aa5c8d}, \href
  {http://adsabs.harvard.edu/abs/2017ApJ...836..230C} {836, 230}

\bibitem[\protect\citeauthoryear{Cowan, Thielemann  \& Truran}{Cowan
  et~al.}{1991}]{Cowan:1991ca}
Cowan J.~J.,  Thielemann F.-K.,   Truran J.~W.,  1991, Physics Reports, 208,
  267

\bibitem[\protect\citeauthoryear{{Dominik}, {Belczynski}, {Fryer}, {Holz},
  {Berti}, {Bulik}, {Mandel}  \& {O'Shaughnessy}}{{Dominik}
  et~al.}{2012}]{2012ApJ...759...52D}
{Dominik} M.,  {Belczynski} K.,  {Fryer} C.,  {Holz} D.~E.,  {Berti} E.,
  {Bulik} T.,  {Mandel} I.,   {O'Shaughnessy} R.,  2012, \mn@doi [\apj]
  {10.1088/0004-637X/759/1/52}, \href
  {http://adsabs.harvard.edu/abs/2012ApJ...759...52D} {759, 52}

\bibitem[\protect\citeauthoryear{{Dutton} \& {Macci{\`o}}}{{Dutton} \&
  {Macci{\`o}}}{2014}]{2014MNRAS.441.3359D}
{Dutton} A.~A.,  {Macci{\`o}} A.~V.,  2014, \mn@doi [\mnras]
  {10.1093/mnras/stu742}, \href
  {http://adsabs.harvard.edu/abs/2014MNRAS.441.3359D} {441, 3359}

\bibitem[\protect\citeauthoryear{Fischer, Mart{\'\i}nez-Pinedo, Hempel  \&
  Liebend{\"o}rfer}{Fischer et~al.}{2012}]{Fischer:2012cl}
Fischer T.,  Mart{\'\i}nez-Pinedo G.,  Hempel M.,   Liebend{\"o}rfer M.,  2012,
  Physical Review D, 85, 857

\bibitem[\protect\citeauthoryear{{Fong} \& {Berger}}{{Fong} \&
  {Berger}}{2013}]{2013ApJ...776...18F}
{Fong} W.,  {Berger} E.,  2013, \mn@doi [\apj] {10.1088/0004-637X/776/1/18},
  \href {http://adsabs.harvard.edu/abs/2013ApJ...776...18F} {776, 18}

\bibitem[\protect\citeauthoryear{Frebel \& Bromm}{Frebel \&
  Bromm}{2012}]{Frebel:2012ja}
Frebel A.,  Bromm V.,  2012, The Astrophysical Journal, 759, 115

\bibitem[\protect\citeauthoryear{{Fryer}, {Burrows}  \& {Benz}}{{Fryer}
  et~al.}{1998}]{1998ApJ...496..333F}
{Fryer} C.,  {Burrows} A.,   {Benz} W.,  1998, \mn@doi [\apj] {10.1086/305348},
  \href {http://adsabs.harvard.edu/abs/1998ApJ...496..333F} {496, 333}

\bibitem[\protect\citeauthoryear{{Fryer}, {Woosley}  \& {Hartmann}}{{Fryer}
  et~al.}{1999}]{1999ApJ...526..152F}
{Fryer} C.~L.,  {Woosley} S.~E.,   {Hartmann} D.~H.,  1999, \mn@doi [\apj]
  {10.1086/307992}, \href {http://adsabs.harvard.edu/abs/1999ApJ...526..152F}
  {526, 152}

\bibitem[\protect\citeauthoryear{Goriely, Bauswein  \& Janka}{Goriely
  et~al.}{2011}]{Goriely:2011fa}
Goriely S.,  Bauswein A.,   Janka H.-T.,  2011, The Astrophysical Journal
  Letters, 738, L32

\bibitem[\protect\citeauthoryear{Goriely, Bauswein, Just, Pllumbi  \&
  Janka}{Goriely et~al.}{2015}]{Goriely:2015hn}
Goriely S.,  Bauswein A.,  Just O.,  Pllumbi E.,   Janka H.~T.,  2015, Monthly
  Notices of the Royal Astronomical Society, 452, 3894

\bibitem[\protect\citeauthoryear{{Hansen} \& {Phinney}}{{Hansen} \&
  {Phinney}}{1997}]{1997MNRAS.291..569H}
{Hansen} B.~M.~S.,  {Phinney} E.~S.,  1997, \mn@doi [\mnras]
  {10.1093/mnras/291.3.569}, \href
  {http://adsabs.harvard.edu/abs/1997MNRAS.291..569H} {291, 569}

\bibitem[\protect\citeauthoryear{{Hirai}, {Ishimaru}, {Saitoh}, {Fujii},
  {Hidaka}  \& {Kajino}}{{Hirai} et~al.}{2015}]{2015ApJ...814...41H}
{Hirai} Y.,  {Ishimaru} Y.,  {Saitoh} T.~R.,  {Fujii} M.~S.,  {Hidaka} J.,
  {Kajino} T.,  2015, \mn@doi [\apj] {10.1088/0004-637X/814/1/41}, \href
  {http://adsabs.harvard.edu/abs/2015ApJ...814...41H} {814, 41}

\bibitem[\protect\citeauthoryear{{Hirai}, {Ishimaru}, {Saitoh}, {Fujii},
  {Hidaka}  \& {Kajino}}{{Hirai} et~al.}{2017}]{2017MNRAS.466.2474H}
{Hirai} Y.,  {Ishimaru} Y.,  {Saitoh} T.~R.,  {Fujii} M.~S.,  {Hidaka} J.,
  {Kajino} T.,  2017, \mn@doi [\mnras] {10.1093/mnras/stw3342}, \href
  {http://adsabs.harvard.edu/abs/2017MNRAS.466.2474H} {466, 2474}

\bibitem[\protect\citeauthoryear{{Hobbs}, {Lorimer}, {Lyne}  \&
  {Kramer}}{{Hobbs} et~al.}{2005}]{2005MNRAS.360..974H}
{Hobbs} G.,  {Lorimer} D.~R.,  {Lyne} A.~G.,   {Kramer} M.,  2005, \mn@doi
  [\mnras] {10.1111/j.1365-2966.2005.09087.x}, \href
  {http://adsabs.harvard.edu/abs/2005MNRAS.360..974H} {360, 974}

\bibitem[\protect\citeauthoryear{{Ishimaru}, {Wanajo}  \&
  {Prantzos}}{{Ishimaru} et~al.}{2015}]{2015ApJ...804L..35I}
{Ishimaru} Y.,  {Wanajo} S.,   {Prantzos} N.,  2015, \mn@doi [\apjl]
  {10.1088/2041-8205/804/2/L35}, \href
  {http://adsabs.harvard.edu/abs/2015ApJ...804L..35I} {804, L35}

\bibitem[\protect\citeauthoryear{{Kalogera} \& {Lorimer}}{{Kalogera} \&
  {Lorimer}}{2000}]{2000ApJ...530..890K}
{Kalogera} V.,  {Lorimer} D.~R.,  2000, \mn@doi [\apj] {10.1086/308417}, \href
  {http://adsabs.harvard.edu/abs/2000ApJ...530..890K} {530, 890}

\bibitem[\protect\citeauthoryear{{Kalogera} et~al.,}{{Kalogera}
  et~al.}{2004}]{2004ApJ...614L.137K}
{Kalogera} V.,  et~al., 2004, \mn@doi [\apjl] {10.1086/425868}, \href
  {http://adsabs.harvard.edu/abs/2004ApJ...614L.137K} {614, L137}

\bibitem[\protect\citeauthoryear{Kelley, Ramirez-Ruiz, Zemp, Diemand  \&
  Mandel}{Kelley et~al.}{2010}]{Kelley:2010iq}
Kelley L.~Z.,  Ramirez-Ruiz E.,  Zemp M.,  Diemand J.,   Mandel I.,  2010, The
  Astrophysical Journal Letters, 725, L91

\bibitem[\protect\citeauthoryear{{Klypin}, {Trujillo-Gomez}  \&
  {Primack}}{{Klypin} et~al.}{2011}]{2011ApJ...740..102K}
{Klypin} A.~A.,  {Trujillo-Gomez} S.,   {Primack} J.,  2011, \mn@doi [\apj]
  {10.1088/0004-637X/740/2/102}, \href
  {http://adsabs.harvard.edu/abs/2011ApJ...740..102K} {740, 102}

\bibitem[\protect\citeauthoryear{{Komiya} \& {Shigeyama}}{{Komiya} \&
  {Shigeyama}}{2016}]{2016ApJ...830...76K}
{Komiya} Y.,  {Shigeyama} T.,  2016, \mn@doi [\apj]
  {10.3847/0004-637X/830/2/76}, \href
  {http://adsabs.harvard.edu/abs/2016ApJ...830...76K} {830, 76}

\bibitem[\protect\citeauthoryear{Korobkin, Rosswog, Arcones  \&
  Winteler}{Korobkin et~al.}{2012}]{Korobkin:2012cp}
Korobkin O.,  Rosswog S.,  Arcones A.,   Winteler C.,  2012, Monthly Notices of
  the Royal Astronomical Society, 426, 1940

\bibitem[\protect\citeauthoryear{{Kravtsov}}{{Kravtsov}}{2013}]{2013ApJ...764L..31K}
{Kravtsov} A.~V.,  2013, \mn@doi [\apjl] {10.1088/2041-8205/764/2/L31}, \href
  {http://adsabs.harvard.edu/abs/2013ApJ...764L..31K} {764, L31}

\bibitem[\protect\citeauthoryear{{Lorimer}}{{Lorimer}}{2008}]{2008LRR....11....8L}
{Lorimer} D.~R.,  2008, \mn@doi [Living Reviews in Relativity]
  {10.12942/lrr-2008-8}, \href
  {http://adsabs.harvard.edu/abs/2008LRR....11....8L} {11}

\bibitem[\protect\citeauthoryear{Mart{\'\i}nez-Pinedo, Fischer, Lohs  \&
  Huther}{Mart{\'\i}nez-Pinedo et~al.}{2012}]{MartinezPinedo:2012ev}
Mart{\'\i}nez-Pinedo G.,  Fischer T.,  Lohs A.,   Huther L.,  2012, Physical
  Review Letters, 109, 251104

\bibitem[\protect\citeauthoryear{{Matteucci}, {Romano}, {Arcones}, {Korobkin}
  \& {Rosswog}}{{Matteucci} et~al.}{2014}]{2014MNRAS.438.2177M}
{Matteucci} F.,  {Romano} D.,  {Arcones} A.,  {Korobkin} O.,   {Rosswog} S.,
  2014, \mn@doi [\mnras] {10.1093/mnras/stt2350}, \href
  {http://adsabs.harvard.edu/abs/2014MNRAS.438.2177M} {438, 2177}

\bibitem[\protect\citeauthoryear{{Mennekens} \& {Vanbeveren}}{{Mennekens} \&
  {Vanbeveren}}{2016}]{2016A&A...589A..64M}
{Mennekens} N.,  {Vanbeveren} D.,  2016, \mn@doi [\aap]
  {10.1051/0004-6361/201628193}, \href
  {http://adsabs.harvard.edu/abs/2016A%26A...589A..64M} {589, A64}

\bibitem[\protect\citeauthoryear{{Moster}, {Naab}  \& {White}}{{Moster}
  et~al.}{2013}]{2013MNRAS.428.3121M}
{Moster} B.~P.,  {Naab} T.,   {White} S.~D.~M.,  2013, \mn@doi [\mnras]
  {10.1093/mnras/sts261}, \href
  {http://adsabs.harvard.edu/abs/2013MNRAS.428.3121M} {428, 3121}

\bibitem[\protect\citeauthoryear{Nishimura, Takiwaki  \& Thielemann}{Nishimura
  et~al.}{2015}]{Nishimura:2015ki}
Nishimura N.,  Takiwaki T.,   Thielemann F.-K.,  2015, The Astrophysical
  Journal, 810, 109

\bibitem[\protect\citeauthoryear{Oechslin, Rosswog  \& Thielemann}{Oechslin
  et~al.}{2002}]{Oechslin:2002iu}
Oechslin R.,  Rosswog S.,   Thielemann F.-K.,  2002, Physical Review D, 65,
  103005

\bibitem[\protect\citeauthoryear{Piran, Nakar  \& Rosswog}{Piran
  et~al.}{2013}]{Piran:2013da}
Piran T.,  Nakar E.,   Rosswog S.,  2013, Monthly Notices of the Royal
  Astronomical Society, 430, 2121

\bibitem[\protect\citeauthoryear{{Planck Collaboration} et~al.,}{{Planck
  Collaboration} et~al.}{2016}]{2016A&A...594A..13P}
{Planck Collaboration} et~al., 2016, \mn@doi [\aap]
  {10.1051/0004-6361/201525830}, \href
  {http://adsabs.harvard.edu/abs/2016A%26A...594A..13P} {594, A13}

\bibitem[\protect\citeauthoryear{Roberts, Reddy  \& Shen}{Roberts
  et~al.}{2012}]{Roberts:2012gc}
Roberts L.~F.,  Reddy S.,   Shen G.,  2012, Physical Review C, 86, 065803

\bibitem[\protect\citeauthoryear{Rosswog, Liebend{\"o}rfer, Thielemann, Davies,
  Benz  \& Piran}{Rosswog et~al.}{1999}]{Rosswog:1999wz}
Rosswog S.,  Liebend{\"o}rfer M.,  Thielemann F.~K.,  Davies M.~B.,  Benz W.,
  Piran T.,  1999, Astronomy {\&} Astrophysics, 341, 499

\bibitem[\protect\citeauthoryear{Rosswog, Davies, Thielemann  \& Piran}{Rosswog
  et~al.}{2000}]{Rosswog:2000nj}
Rosswog S.,  Davies M.~B.,  Thielemann F.~K.,   Piran T.,  2000, \mnras, 360,
  171

\bibitem[\protect\citeauthoryear{Shappee \& Thompson}{Shappee \&
  Thompson}{2013}]{Shappee:2013ea}
Shappee B.~J.,  Thompson T.~A.,  2013, The Astrophysical Journal, 766, 64

\bibitem[\protect\citeauthoryear{Shen, Cooke, Ramirez-Ruiz, Madau, Mayer  \&
  Guedes}{Shen et~al.}{2015}]{Shen:2015gc}
Shen S.,  Cooke R.~J.,  Ramirez-Ruiz E.,  Madau P.,  Mayer L.,   Guedes J.,
  2015, The Astrophysical Journal, 807, 115

\bibitem[\protect\citeauthoryear{Vargas, Geha, Kirby  \& Simon}{Vargas
  et~al.}{2013}]{Vargas:2013ei}
Vargas L.~C.,  Geha M.,  Kirby E.~N.,   Simon J.~D.,  2013, The Astrophysical
  Journal, 767, 134

\bibitem[\protect\citeauthoryear{Wanajo}{Wanajo}{2013}]{Wanajo:2013io}
Wanajo S.,  2013, The Astrophysical Journal, 770, L22

\bibitem[\protect\citeauthoryear{Wanajo, Janka  \& M~ller}{Wanajo
  et~al.}{2010}]{Wanajo:2010js}
Wanajo S.,  Janka H.-T.,   M~ller B.,  2010, The Astrophysical Journal Letters,
  726, L15

\bibitem[\protect\citeauthoryear{Wanajo, Sekiguchi, Nishimura, Kiuchi, Kyutoku
  \& Shibata}{Wanajo et~al.}{2014}]{Wanajo:2014ia}
Wanajo S.,  Sekiguchi Y.,  Nishimura N.,  Kiuchi K.,  Kyutoku K.,   Shibata M.,
   2014, The Astrophysical Journal Letters, 789, L39

\bibitem[\protect\citeauthoryear{{Wang}, {Han}, {Cooper}, {Cole}, {Frenk}  \&
  {Lowing}}{{Wang} et~al.}{2015}]{2015MNRAS.453..377W}
{Wang} W.,  {Han} J.,  {Cooper} A.~P.,  {Cole} S.,  {Frenk} C.,   {Lowing} B.,
  2015, \mn@doi [\mnras] {10.1093/mnras/stv1647}, \href
  {http://adsabs.harvard.edu/abs/2015MNRAS.453..377W} {453, 377}

\bibitem[\protect\citeauthoryear{{Wehmeyer}, {Pignatari}  \&
  {Thielemann}}{{Wehmeyer} et~al.}{2015}]{2015MNRAS.452.1970W}
{Wehmeyer} B.,  {Pignatari} M.,   {Thielemann} F.-K.,  2015, \mn@doi [\mnras]
  {10.1093/mnras/stv1352}, \href
  {http://adsabs.harvard.edu/abs/2015MNRAS.452.1970W} {452, 1970}

\bibitem[\protect\citeauthoryear{Winteler, K{\"a}ppeli, Perego, Arcones,
  Vasset, Nishimura, Liebend{\"o}rfer  \& Thielemann}{Winteler
  et~al.}{2012}]{Winteler:2012fv}
Winteler C.,  K{\"a}ppeli R.,  Perego A.,  Arcones A.,  Vasset N.,  Nishimura
  N.,  Liebend{\"o}rfer M.,   Thielemann F.~K.,  2012, The Astrophysical
  Journal Letters, 750, L22

\bibitem[\protect\citeauthoryear{Woosley, Wilson, Mathews, Hoffman  \&
  Meyer}{Woosley et~al.}{1994}]{Woosley:1994ih}
Woosley S.~E.,  Wilson J.~R.,  Mathews G.~J.,  Hoffman R.~D.,   Meyer B.~S.,
  1994, Astrophysical Journal, 433, 229

\bibitem[\protect\citeauthoryear{Zemp, Ramirez-Ruiz  \& Diemand}{Zemp
  et~al.}{2009}]{Zemp:2009gu}
Zemp M.,  Ramirez-Ruiz E.,   Diemand J.,  2009, The Astrophysical Journal, 705,
  L186

\bibitem[\protect\citeauthoryear{van~de Voort, Quataert, Hopkins, Kere{\v s}
  \& Faucher-Gigu{\`e}re}{van~de Voort et~al.}{2015}]{vandeVoort:2015jw}
van~de Voort F.,  Quataert E.,  Hopkins P.~F.,  Kere{\v s} D.,
  Faucher-Gigu{\`e}re C.-A.,  2015, Monthly Notices of the Royal Astronomical
  Society, 447, 140

\bibitem[\protect\citeauthoryear{{van der Wel} et~al.,}{{van der Wel}
  et~al.}{2014}]{2014ApJ...788...28V}
{van der Wel} A.,  et~al., 2014, \mn@doi [\apj] {10.1088/0004-637X/788/1/28},
  \href {http://adsabs.harvard.edu/abs/2014ApJ...788...28V} {788, 28}

\makeatother
\end{thebibliography}
%\expandafter\ifx\csname %natexlab\endcsname\relax\def\natexlab#1{#1}\fi
\bsp
\label{lastpage}

%\end{thebibliography}

\end{document}